\documentclass[sigconf,authorversion, nonacm]{acmart}

\setcopyright{none}
\AtBeginDocument{%
  }

\usepackage{booktabs}
\usepackage{multirow}
\usepackage{amsmath}
\usepackage{array}
\usepackage{float}
\usepackage{graphicx}
\usepackage{hyperref}
\graphicspath{{./}{workshop/}}
\newcolumntype{L}[1]{>{\raggedright\arraybackslash}p{#1}}

\begin{document}

\title{STAR: Structured Tokenization and Target-Aware Interest Representation for PCVR Prediction}

\author{Yimeng Xu}
\authornote{These authors contributed equally to this work.}
\affiliation{%
  \institution{Tsinghua University}
  \city{Shenzhen}
  \country{China}}
\email{xuym24@mails.tsinghua.edu.cn}

\author{Ruihao Zhang}
\authornotemark[1]
\affiliation{%
  \institution{Peking University}
  \city{Beijing}
  \country{China}}
\email{2401210530@stu.pku.edu.cn}

\author{Yingqi Song}
\authornotemark[1]
\affiliation{%
  \institution{Tsinghua University}
  \city{Shenzhen}
  \country{China}}
\email{songyq24@mails.tsinghua.edu.cn}

\author{Ying Jiang}
\affiliation{%
  \institution{Tsinghua University}
  \city{Shenzhen}
  \country{China}}
\email{y-jiang25@mails.tsinghua.edu.cn}

\author{Lan Ma}
\authornote{Correponding author.}
\affiliation{%
  \institution{Tsinghua University}
  \city{Shenzhen}
  \country{China}}
\email{malan@sz.tsinghua.edu.cn}
\renewcommand{\shortauthors}{Xu, Zhang, Song, Jiang, and Ma}

\begin{abstract}
Post-click conversion rate (PCVR) prediction is a core ranking task
in industrial recommender systems.
Modern ranking models must jointly capture heterogeneous non-sequential
features, multi-behavior user sequences, and target-item-aware user
interests, while remaining robust to
high-cardinality sparse features, missing values, and train-inference
inconsistencies. In this paper, we present STAR (Structured
Tokenization and Target-Aware Interest Representation), a practical
framework for the KDD Cup 2026 Tencent UniRec Challenge. STAR combines
structured feature tokenization with target-aware interest representation
on top of a HyFormer-style multi-sequence backbone. It introduces
high-cardinality signal recovery, explicit user-item interaction tokens,
target-aware sequence decoding, and a weighted user-item contrastive
auxiliary objective inspired by InfoNCE.
We further align the training and inference pipelines by reconstructing
feature remapping tables and structural hyperparameters from the saved
training configuration. Experiments on the challenge dataset identify
the components that most reliably improve ranking AUC, while LogLoss is
reported as a calibration diagnostic. The main ablation study shows a
large gain from temporal context, with smaller but useful contributions
from contrastive alignment, target-aware interest encoding, and
high-cardinality sequence feature recovery. The source code is publicly available \href{https://github.com/AIzealotwu/taac_26_academic_rank2_firstround_rank11_secondround}{in our GitHub repository}..
\end{abstract}


\maketitle

\section{Introduction}

Post-click conversion rate (PCVR) prediction estimates the probability
that a user converts after interacting with a candidate item, and is
typically formulated as an entire-space or post-click ranking
problem~\cite{ma2018esmm}. It is a challenging industrial ranking task
because the model must combine heterogeneous non-sequential features,
sparse user/item identifiers, dense numerical signals, and multiple
user behavior sequences~\cite{covington2016youtube,cheng2016widedeep}.

Early deep ranking models focus on explicit and implicit feature
interactions, e.g., factorization machines~\cite{rendle2010fm},
Wide\&Deep~\cite{cheng2016widedeep},
DeepFM~\cite{guo2017deepfm}, and
DCN/DCN-V2~\cite{wang2017dcn,wang2021dcnv2}. On the sequence side, a
separate line of work models user behavior histories with target
attention~\cite{zhou2018din,zhou2019dien}, self-attention
sequence encoders~\cite{kang2018sasrec,sun2019bert4rec,chen2019behavior},
and search-based long-history mechanisms~\cite{pi2020sim,lyu2025dv365}.
More recent industrial ranking architectures further scale up ranking
models~\cite{zhu2026rankmixer,chen2026rankup} and unify
feature interaction with behavior sequence modeling in a single
Transformer-style backbone, exemplified by
HyFormer~\cite{huang2026hyformer}, MixFormer~\cite{huang2026mixformer},
and OneTrans~\cite{zhang2026onetrans}. However, in practical PCVR
systems, we observe three additional challenges. First,
high-cardinality sparse fields are often skipped because their embedding
tables are too expensive, causing useful long-tail signals to be lost,
as also observed in ultra-long history
modeling~\cite{lyu2025dv365}. Second, user
sequence representations need to be conditioned on the candidate item,
otherwise the extracted interest representation may be insufficiently
target-aware~\cite{zhou2018din}. Third, missing values, padding zeros,
and feature distribution shifts can create train-inference mismatch and
unstable optimization.

To address these issues, we build STAR, a framework that combines
structured tokenization and target-aware interest representation on top
of the provided formal baseline~\cite{huang2026hyformer}. Our approach
augments the baseline with high-cardinality feature recovery,
target-aware sequence decoding, user-item interaction tokens, robust
missing-value handling, and an auxiliary InfoNCE-style alignment
objective~\cite{oord2018cpc,radford2021clip}.

The main contributions are summarized as follows:
\begin{itemize}
  \item We design a structured feature tokenization pipeline that
  organizes sparse IDs, dense numerical fields, temporal context, and
  sequence-derived signals as tokens. It also recovers high-cardinality
  sparse signals using cardinality capping, frequency remapping, and
  sequence hashing, while explicitly modeling missing values and padding
  zeros.
  \item We evaluate target-aware interest-representation enhancements,
  including DIN-style~\cite{zhou2018din} query decoding, interest
  tokens, aligned sparse-dense pair pooling, and user-item
  non-sequential pair-product tokens.
  \item We introduce a weighted user-item contrastive objective
  inspired by InfoNCE~\cite{oord2018cpc,radford2021clip} and a
  training-inference consistency mechanism that reconstructs remapping
  tables and model structure during inference.
\end{itemize}

\section{Related Work}

\subsection{Feature Interaction Models}

Feature interaction is central to CTR and CVR prediction. Classical
models use factorization~\cite{rendle2010fm} or cross
networks~\cite{wang2017dcn,wang2021dcnv2} to model low- and high-order
interactions among sparse and dense fields. Hybrid architectures such
as Wide\&Deep~\cite{cheng2016widedeep}, DeepFM~\cite{guo2017deepfm},
xDeepFM~\cite{lian2018xdeepfm}, and
AutoInt~\cite{song2019autoint} combine memorization and generalization
by jointly training explicit and implicit interaction branches. More
recent industrial systems adopt token-mixing or Transformer-style
modules~\cite{vaswani2017attention} to improve expressiveness under
large-scale sparse feature inputs and to better exploit modern
GPUs~\cite{zhu2026rankmixer,chen2026rankup}.

\subsection{Sequence Modeling for Recommendation}

User behavior sequences provide important signals for preference
modeling. Target attention and DIN-style activation
units~\cite{zhou2018din,zhou2019dien}, self-attention sequence
encoders~\cite{kang2018sasrec,sun2019bert4rec}, and behavior sequence
Transformers~\cite{chen2019behavior} have been widely used to extract
user interests from historical actions. To scale to ultra-long user
histories under strict latency budgets, recent works explore
search-based selection~\cite{pi2020sim} and extremely long history
modeling at web scale~\cite{lyu2025dv365}. In PCVR
prediction~\cite{ma2018esmm}, target-aware sequence modeling is
especially important because the relevant subset of history depends on
the candidate item.

\subsection{Unified Sequence-Feature Architectures}

Recent industrial ranking models such as
HyFormer~\cite{huang2026hyformer},
MixFormer~\cite{huang2026mixformer}, and
OneTrans~\cite{zhang2026onetrans} seek to jointly model non-sequential
features and behavior sequences within a single Transformer-style
backbone. They typically tokenize heterogeneous inputs and alternate
between sequence modeling and feature
interaction~\cite{vaswani2017attention}, and are complementary to
scaling-oriented ranking backbones such as
RankMixer~\cite{zhu2026rankmixer} and
RankUp~\cite{chen2026rankup}. Our work follows this unified
sequence-feature line but focuses on practical robustness for the
UniRec PCVR setting, including high-cardinality feature recovery,
target-aware query decoding~\cite{zhou2018din}, and a user-item
contrastive auxiliary objective~\cite{oord2018cpc,radford2021clip}.

\section{Methodology}
\begin{figure*}[t]
  \centering
  \includegraphics[width=\textwidth]{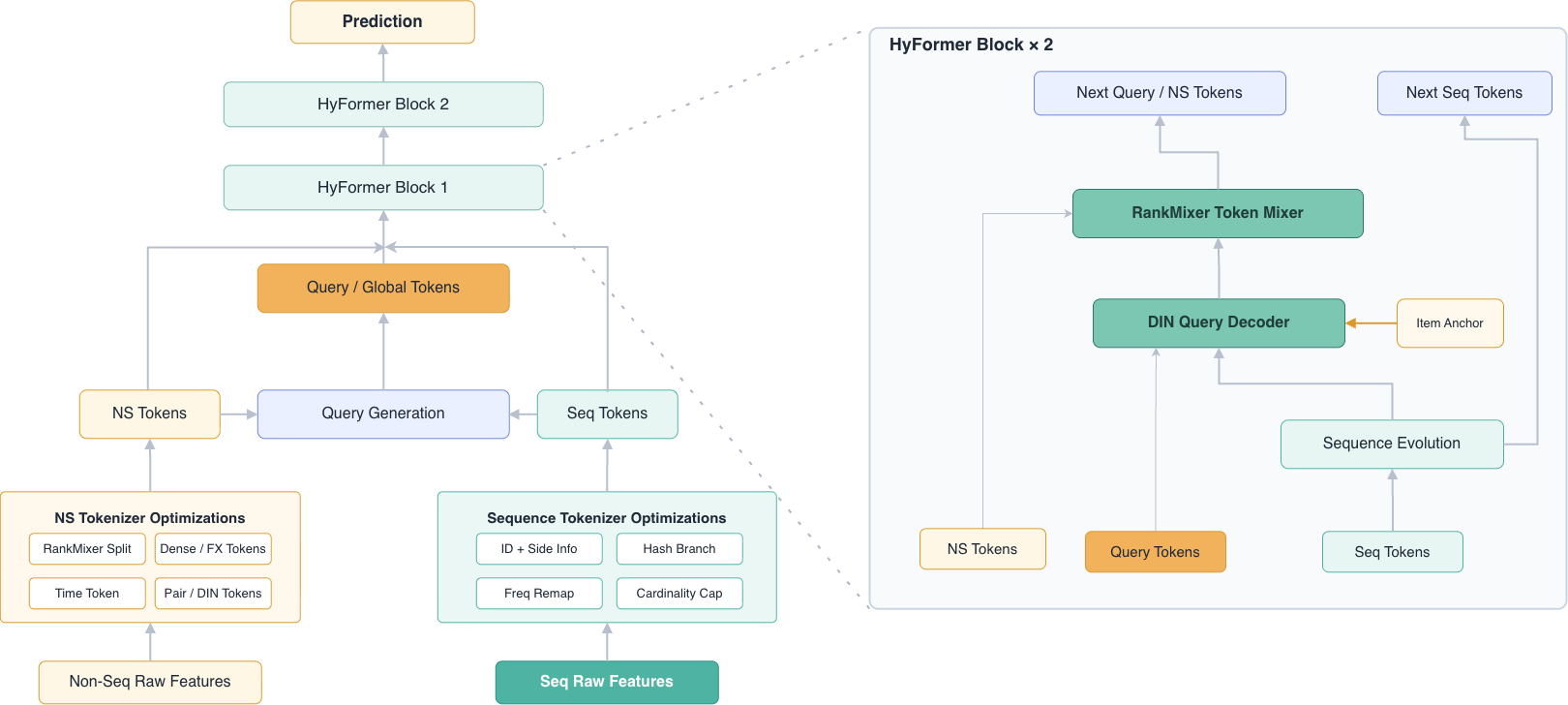}
  \caption{Overall architecture of STAR.}
  \Description{The framework first applies optimized tokenizers to
  non-sequential and sequence features, then generates query/global
  tokens and processes them with target-aware sequence-feature blocks before
  producing the PCVR prediction.}
  \label{fig:framework}
\end{figure*}

\subsection{Problem Formulation}

Let $\mathcal{U}$ and $\mathcal{I}$ denote the user and item spaces.
For a user $u \in \mathcal{U}$ and a candidate item $i \in
\mathcal{I}$, we observe non-sequential features $\mathbf{x}_{u,i}$
and a set of behavior sequences $\mathcal{S}_u =
\{S_u^{(1)}, \ldots, S_u^{(K)}\}$. The goal is to predict the
post-click conversion probability
\begin{equation}
  \hat{y}_{u,i} = f_{\theta}(\mathbf{x}_{u,i}, \mathcal{S}_u),
\end{equation}
where $y_{u,i} \in \{0,1\}$ is the conversion label. 
The primary training loss is binary cross-entropy (BCE) or its focal loss variant. For standard BCE, the loss is:
\begin{equation}
  \mathcal{L}_{\mathrm{bce}} = - y \log \hat{y} - (1-y)\log(1-\hat{y}).
\end{equation}
To address extreme class imbalance, the system defaults to using a focal loss, which introduces a modulating factor to down-weight easy examples and a class-balancing term $\alpha$:
\begin{equation}
  \mathcal{L}_{\mathrm{focal}} = - \alpha_t (1 - p_t)^\gamma \log(p_t),
\end{equation}
where $p_t = \hat{y}$ if $y=1$ and $1-\hat{y}$ otherwise, and $\alpha_t = \alpha$ if $y=1$ and $1-\alpha$ otherwise. We set the focusing parameter $\gamma=2.0$ and the positive-class weight $\alpha=0.366$ in our default configuration.

\subsection{Overall Framework}

Figure~\ref{fig:framework} illustrates STAR. The system contains four components: structured feature tokenization, target-aware interest representation, user-item interaction enhancement, and an auxiliary contrastive training objective.

\subsection{Structured Feature Tokenization}

\begin{figure}
    \centering
    \IfFileExists{tokenizer_detail_gemini_bai_final_version.png}{%
      \includegraphics[width=\linewidth]{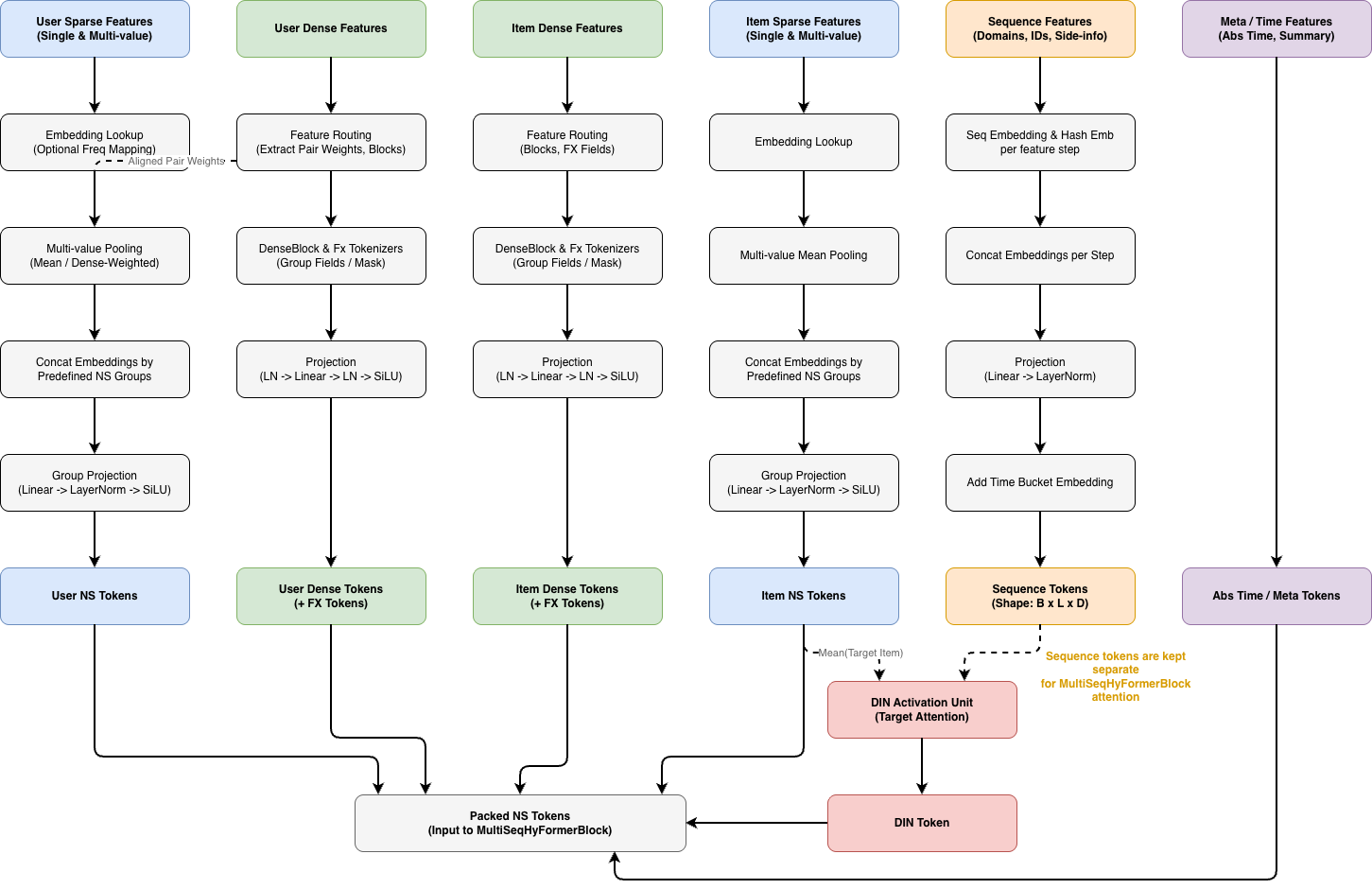}%
    }{%
      \fbox{\parbox[c][0.36\linewidth][c]{0.9\linewidth}{\centering
      Structured tokenization pipeline figure placeholder}}%
    }
    \caption{Structured tokenization pipeline of STAR.}
    \label{fig:placeholder}
\end{figure}

\paragraph{Cardinality capping.}

For ultra-high-cardinality user integer fields, a full embedding table
is impractical. We keep the most frequent IDs as dedicated embedding
rows and fold the rare tail into a shared bucket. In this way, both specificity and commonality are effectively learned within the high-frequency IDs and the unknown category, respectively. This allows fields that would otherwise be skipped to re-enter the model with a compact trainable representation.

\paragraph{Frequency remapping and sequence hashing.}

For selected high-cardinality fields, frequent IDs are remapped to dedicated compact indices, whereas rare or unseen IDs share an unknown bucket. 
For sequence features that exceed the embedding threshold, we optionally route IDs through a fixed-size hash embedding branch. Thus, the features can be effectively utilized while preserving the specificity of the high-frequency features.

\paragraph{Missing-value and padding-zero disambiguation.}

Raw feature value zero can either mean a true value or padding. We therefore add non-zero masks for dense features and shift integer IDs when enabled so that padding remains index zero. 
Empirically, we implement zero-variance feature pruning. For fields with no variance across the dataset, instead of embedding their raw IDs, we explicitly drop their values and only extract their missing indicators, which are then projected into the model as residual dense signals.
In addition, dropped features and item dense fields can emit missing indicators that are projected into the model as residual signals.

\paragraph{Time and meta-summary features.}

The model uses absolute target-time features and optional sequence periodic side information. For absolute time features at the sample-level, we select the time of day and whether on a weekend. We also compute sample-level meta summaries, such as sequence lengths, truncation ratios, recency spans, and sentinel-value statistics as our important features, and inject them into the late prediction path and query generation path.

\paragraph{Dense feature tokenization.}

As shown in Figure 2, dense features are treated as semantic continuous fields instead of being naively concatenated in our carefully designed feature pipeline. Raw Parquet float-list columns are first padded or truncated according to the schema, with missing or empty values represented by zeros. To separate real zero values from padding, selected dense blocks append a non-zero indicator mask. We then tokenize dense fields by their semantics: aligned sparse-dense pairs are used as weights for sparse multi-hot pooling, while remaining dense fields are projected into dedicated dense tokens or field-specific auxiliary tokens. These tokens are concatenated with user/item non-sequential tokens and sequence-derived query tokens for the final sequence-feature interaction.

\subsection{Target-Aware Interest Representation}

As shown in the left part of  Figure ~\ref{fig:framework}, the backbone follows a multi-sequence HyFormer design. Non-sequential features are tokenized into user, item, dense, and auxiliary tokens.
Each behavior domain is embedded into a sequence-token matrix. A query generator produces sequence-specific query tokens conditioned on non-sequential tokens and pooled sequence summaries.
Unlike the formal baseline, our block uses a DIN-style query decoder. Given a sequence-specific query token $\mathbf{q}$, an item anchor $\mathbf{h}_i$, and sequence tokens $\mathbf{h}_t$, the decoder first generates a target representation $\mathbf{q}_{\text{tgt}}$ and computes attention weights $\alpha_t$:
\begin{equation}
    \mathbf{q}_{\text{tgt}} = \text{MLP}_{\text{target}}([\mathbf{q}, \mathbf{h}_i, \mathbf{q} - \mathbf{h}_i, \mathbf{q} \odot \mathbf{h}_i])
\end{equation}
\begin{equation}
    \alpha_t = \text{softmax} (\text{MLP}_{\text{score}}([\mathbf{h}_t, \mathbf{q}_{\text{tgt}}, \mathbf{h}_t - \mathbf{q}_{\text{tgt}}, \mathbf{h}_t \odot \mathbf{q}_{\text{tgt}}]))
\end{equation}
The pooled interest vector $\mathbf{z} = \sum_t \alpha_t \mathbf{h}_t$ is then used to update the query via a residual connection:
\begin{equation}
    \mathbf{q}_{\text{out}} = \mathbf{q} + \text{MLP}_{\text{update}}([\mathbf{q}, \mathbf{z}, \mathbf{h}_i, \mathbf{q} \odot \mathbf{z}, \mathbf{h}_i \odot \mathbf{z}])
\end{equation}

\subsection{User-Item Interaction Enhancement}

We add two interaction mechanisms. 
First, aligned sparse-dense pair pooling uses dense values as weights over matching sparse embeddings for selected feature IDs. This avoids modeling the same signal twice and gives the model a direct weighted representation of positive pair features. 
Second, user and item non-sequential tokens are combined with element-wise pair-product projections, yielding explicit user-item interaction tokens. 
Furthermore, before feeding tokens into the HyFormer blocks, we compute an explicit target-aware interest token. Specifically, we apply a DIN activation unit over the raw sequence embeddings $\mathbf{h}_t$ anchored by the candidate item $\mathbf{h}_i$ to compute attention weights and an interest vector $\mathbf{z}$:
\begin{equation}
    \alpha_t = \text{softmax} (g([\mathbf{h}_i, \mathbf{h}_t, \mathbf{h}_i - \mathbf{h}_t, \mathbf{h}_i \odot \mathbf{h}_t])) , \quad \mathbf{z} = \sum_t \alpha_t \mathbf{h}_t
\end{equation}
where $g(\cdot)$ is a multi-layer perceptron. We then fuse the outputs $\mathbf{z}$ across all behavior domains via an MLP, and append this fused vector as an additional non-sequential token.

\subsection{Auxiliary User-Item Contrastive Objective}

To align user and item representations, we project pooled user and item tokens into a shared contrastive space. For a mini-batch, the matching user-item pair is treated as the positive pair, while other pairs act as in-batch negatives. The symmetric InfoNCE objective is
\begin{equation}
  \mathcal{L}_{\mathrm{nce}} =
  \frac{1}{2}\left[
  \mathrm{CE}\left(\frac{\mathbf{U}\mathbf{V}^{\top}}{\tau}, \mathbf{t}\right)
  +
  \mathrm{CE}\left(\frac{\mathbf{V}\mathbf{U}^{\top}}{\tau}, \mathbf{t}\right)
  \right],
\end{equation}
where $\tau$ is the temperature and $\mathbf{t}$ indexes matching pairs.
Converted and non-converted samples use different pair weights. The final objective is
\begin{equation}
  \mathcal{L} =
  \mathcal{L}_{\mathrm{main}}
  + \lambda(t) \mathcal{L}_{\mathrm{nce}},
\end{equation}
where $\lambda(t)$ decays during training and is capped by the main loss ratio for stability.

\subsection{Training and Inference Consistency}

By using global negative sample pool, the training pipeline supports fixed global-batch manifests so that runs with different GPU counts can consume equivalent global sample sets. 
Regarding the data shuffle strategy, to ensure robust optimization on sequential log data, we employ a two-level buffer-based approach.  We first globally shuffle the read order of data row groups, and then maintain an in-memory buffer spanning multiple batches to perform local row-level permutations. 
As for optimization tricks, we also use bf16 automatic mixed precision, gradient
accumulation, dense-parameter EMA, and strict checkpoint sidecar files. We serialize training configurations and feature schemas during training to ensure strict alignment.
Train-Inference Alignment, during inference, the dataset and model are reconstructed from the training configuration file; cardinality-cap tables and frequency mapping tables are reloaded so that the embedding layout exactly matches the trained checkpoint.

\section{Experiments}

We focus the experimental section on controlled single-run studies of the optimized system. 
We first evaluate how the model scales with hidden
width, embedding dimension, and the number of HyFormer blocks, and then isolate individual modeling components through ablations. The goal is to separate capacity effects from component-level evidence and to identify which changes should be treated as optional or requiring multi-seed confirmation. 
Unless otherwise stated, the tables below use the full
optimized default-capacity model as the reference.

\subsection{Experimental Setting}

\paragraph{Dataset and evaluation protocol.}
We use the KDD Cup 2026 Tencent UniRec Challenge PCVR dataset with 20000000 training samples. Each
sample contains anonymized user and item identifiers, non-sequential integer and dense features, four behavior-sequence domains, a target
timestamp, and a binary conversion label derived from the official label
type. Unless otherwise specified, all variants use the same row-group
held-out validation split and the same sequence truncation lengths
(\texttt{seq\_a}/\texttt{seq\_b}: 256 and
\texttt{seq\_c}/\texttt{seq\_d}: 512). The training, validation, and
inference scripts rebuild feature schemas from the same configuration so
that each ablation is evaluated with a consistent field order and
embedding layout.

\paragraph{Metrics and selection.}
We report validation AUC, validation LogLoss, and test AUC. AUC is the
primary metric because the challenge is a ranking task; LogLoss is used
as a calibration diagnostic and secondary stability signal. Checkpoints
are selected by validation AUC. For all ablations, $\Delta$AUC is
computed against the full optimized model, so a negative value means that
removing the component hurts performance.

\paragraph{Implementation details.}
For the full system, we set $d_{\mathrm{model}}=128$, sparse embedding
dimension to $64$, the number of HyFormer blocks to $2$, and the global
batch size to $5880$. Training uses focal loss, bf16 mixed precision,
gradient accumulation, dense-parameter EMA, and a fixed global-batch
manifest. The InfoNCE temperature is $0.07$; its weight decays from
$0.1$ to $0.01$ during early training and is capped relative to the main
loss for stability. Unless an ablation explicitly disables a module, all
variants use the same optimizer, split, sequence lengths, and checkpoint
selection rule.

\subsection{Full-System Reference}

Table~\ref{tab:full_reference} compares the official baseline with the
optimized full model used as the reference for all subsequent ablations.
The full model (\texttt{best}) improves validation/test AUC by
0.014921/0.016585, while LogLoss is higher; we therefore keep AUC as the
primary metric and use LogLoss as a calibration diagnostic.
\texttt{best} configuration.

\begin{table}[H]
  \centering
  \caption{Official baseline and full-system reference.}
  \label{tab:full_reference}
  \small
  \setlength{\tabcolsep}{4pt}
  \footnotesize
\setlength{\tabcolsep}{3pt}
\begin{tabular}{@{}lcccc@{}}
  \toprule
  Model & Val AUC $\uparrow$  & Test AUC $\uparrow$ & $\Delta$ Test AUC \\
  \midrule
  Official baseline & 0.829582  & 0.819961 & 0.000000 \\
  Full optimized model & 0.844503  & 0.836546 & +0.016585 \\
  \bottomrule
\end{tabular}
\end{table}

\subsection{Model Scaling}

\begin{figure}[H]
  \centering
  \includegraphics[width=\columnwidth]{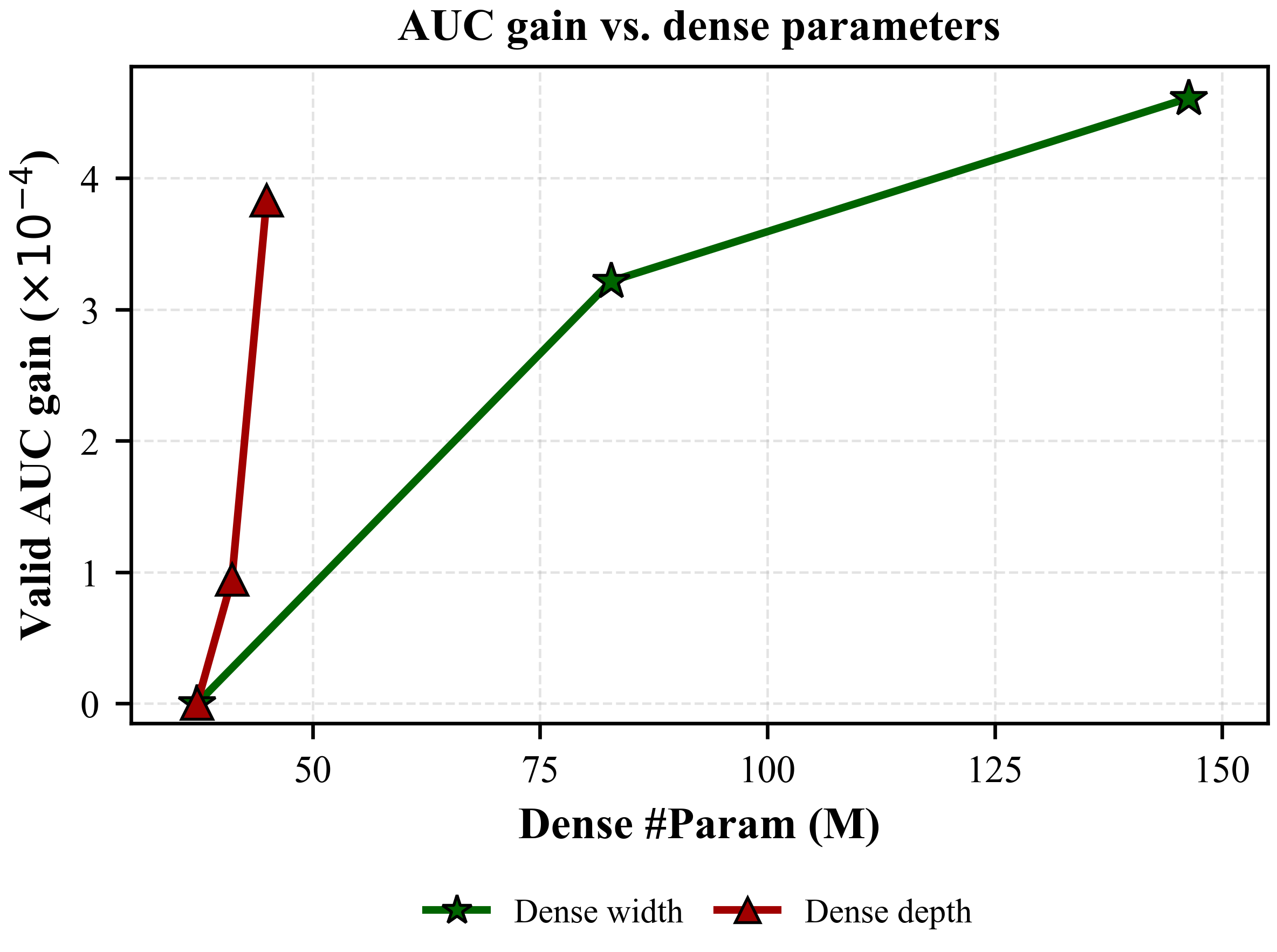}
  \caption{Validation-AUC scaling curves over dense-network parameters.
  The two curves isolate hidden-width scaling and block-depth scaling.
  Gains are measured against the default full model.}
  \label{fig:scaling_curves}
\end{figure}

Table~\ref{tab:model_scaling} studies model capacity while keeping the
same optimized training recipe and feature pipeline.
Figure~\ref{fig:scaling_curves} shows the two clean dense-scaling
directions. Dense-width scaling gives the clearest monotonic validation
gain: increasing $d_{\mathrm{model}}$ from 128 to 192 and 256 raises
validation AUC by 3.22 and 4.61 points in $10^{-4}$ units, with
corresponding test AUC gains of 0.000179 and 0.000219. Depth scaling is
more parameter-efficient: increasing the number of HyFormer blocks from
2 to 4 adds only 7.65M dense parameters and improves test AUC by
0.000199. In contrast, embedding-dimension scaling is less monotonic, so
we discuss it separately in Appendix~\ref{sec:additional_scaling}.

\begin{table*}[!t]
  \centering
  \caption{Model-scaling results. Parameter counts are shown in millions,
  separated into sparse embedding parameters and dense network parameters.
  $\Delta$AUC is computed against the default full optimized model.}
  \label{tab:model_scaling}
  \footnotesize
  \setlength{\tabcolsep}{3pt}
  \begin{tabular}{@{}L{0.18\textwidth}rrrrrrrrr@{}}
    \toprule
    Variant & $d$ & Emb. & Blocks & Sparse (M) & Dense (M)
    & Valid AUC & Valid LogLoss & Test AUC & $\Delta$ Test AUC \\
    \midrule
    Default & 128 & 64 & 2 & 333.97 & 37.28
    & 0.844503 & 0.296215 & 0.836546 & 0.000000 \\
    Wider hidden & 192 & 64 & 2 & 333.98 & 82.82
    & 0.844825 & 0.295331 & 0.836725 & +0.000179 \\
    Widest hidden & 256 & 64 & 2 & 333.98 & 146.29
    & 0.844964 & 0.294054 & 0.836765 & +0.000219 \\
    Wider embedding & 128 & 96 & 2 & 500.96 & 37.79
    & 0.844906 & 0.295928 & 0.836604 & +0.000058 \\
    Hidden+embedding & 192 & 128 & 2 & 667.94 & 84.34
    & 0.844736 & 0.295478 & 0.836771 & +0.000225 \\
    Deeper, 3 blocks & 128 & 64 & 3 & 333.97 & 41.11
    & 0.844598 & 0.295587 & 0.836666 & +0.000120 \\
    Deeper, 4 blocks & 128 & 64 & 4 & 333.97 & 44.93
    & 0.844887 & 0.295164 & 0.836745 & +0.000199 \\
    Large model & 192 & 128 & 4 & 667.94 & 101.53
    & 0.844867 & 0.293685 & 0.836763 & +0.000217 \\
    \bottomrule
  \end{tabular}
\end{table*}

\subsection{Main Modeling Ablations}

Table~\ref{tab:model_ablation} keeps only ablations whose removal lowers
test AUC. This makes the main table focus on components that are
supported by the current evidence, while mixed or score-increasing
ablations are reported in the appendix.
Temporal context is the dominant modeling contributor: removing absolute
time features reduces test AUC by 0.002043. InfoNCE contributes the
largest remaining modeling gain, suggesting that user-item representation
alignment helps ranking even though LogLoss is not improved. DIN-based
query decoding and high-cardinality sequence recovery provide smaller but
consistent gains. The explicit user-item pair-product token is included
because its removal is negative on test AUC, but the effect is marginal
and should be confirmed with additional seeds.

\begin{table*}[!t]
  \centering
  \caption{Main modeling ablations whose removal lowers test AUC. $\Delta$AUC
  is computed against the full optimized model.}
  \label{tab:model_ablation}
  \small
  \setlength{\tabcolsep}{3.5pt}
  \begin{tabular}{@{}L{0.22\textwidth}L{0.35\textwidth}rrrr@{}}
    \toprule
    Variant & Removed or changed component
    & Valid AUC & Valid LogLoss & Test AUC & $\Delta$ Test AUC \\
    \midrule
    Full model & None
    & 0.844503 & 0.296215 & 0.836546 & 0.000000 \\
    w/o time & Absolute target-time features
    & 0.843018 & 0.298041 & 0.834503 & -0.002043 \\
    w/o padding-zero disambiguation & Missing-value and padding-zero disambiguation
    & 0.844606 & 0.295735 & 0.835219 & -0.001327 \\
    w/o InfoNCE & Weighted user-item contrastive objective
    & 0.844214 & 0.295294 & 0.836210 & -0.000336 \\
    w/o DIN decoder & DIN+MLP query decoder replaced by cross-attention
    & 0.844352 & 0.296259 & 0.836353 & -0.000193 \\
    w/o seq hash/freq & Sequence hash embeddings and frequency remapping
    & 0.844429 & 0.296518 & 0.836422 & -0.000124 \\
    w/o UI pair-product & User/item non-sequential pair-product tokens
    & 0.844711 & 0.296397 & 0.836514 & -0.000032 \\
    \bottomrule
  \end{tabular}
\end{table*}

\subsection{Takeaways}

The scaling results show that the optimized architecture benefits from
additional capacity, especially hidden width and depth, but the gains are
modest. The main ablation results support a complementary claim:
temporal context, padding-zero disambiguation, InfoNCE alignment, DIN query decoding, and
high-cardinality sequence recovery are useful modeling changes, while the
pair-product token is a weak positive signal. Other modules and
training-protocol choices are important engineering diagnostics, but we
keep them in the appendix so that the main table does not over-claim
components that are mixed, score-increasing after removal, or outside
the modeling contribution.

\section{Discussion}

The proposed system follows a different emphasis from simply scaling a
larger Transformer backbone. The formal baseline already provides a
reasonable HyFormer-style architecture for mixing non-sequential
features and behavior sequences. However, our empirical analysis reveals that several practical PCVR signals are missing or weakly modeled in that
baseline: ultra-wide fields are skipped, zeros can be confused with
padding or missing values, user histories are only weakly conditioned on
the target item, and feature remapping artifacts are not treated as part
of the checkpointed inference contract.

The model-scaling results make this picture more nuanced. Larger hidden
width and deeper HyFormer stacks improve AUC, so part of the final
system quality comes from capacity. However, scaling alone does not
identify which modeling assumptions matter; the gains from dense scaling
remain modest compared with the improvements from several targeted
components. The ablation results make the component picture more selective. At
the input level, absolute-time features are clearly important, and
high-cardinality sequence recovery provides a smaller positive signal. At
the interaction level, DIN-style query decoding is useful, while the
explicit pair-product token is only marginal. At the objective level,
weighted InfoNCE regularizes user-item representations and improves test
AUC. In contrast, several plausible additions, including the fused DIN
token, pair pooling, feature dropping, and meta context, improve AUC when
removed in the current run. We therefore move those mixed ablations to
the appendix and avoid claiming them as core contributions. 

The LogLoss results also show that ranking and calibration do not move
identically. The appendix shows that BCE
substantially lowers validation LogLoss relative to focal loss, while
focal loss keeps a small AUC advantage. We therefore treat AUC as the
primary challenge metric and LogLoss as a diagnostic for
calibration-sensitive deployments. Focal Loss improves ranking performance (AUC) by weighting hard samples (often the minority class), but this compromises the absolute calibration of predicted probabilities, resulting in higher LogLoss.

The main trade-off is operational complexity. Compact vocabularies,
frequency maps, and hash allowlists introduce sidecar artifacts that must
be versioned with the checkpoint. The fixed global-batch manifest and
strict inference reconstruction mitigate this risk, but they also make
the training recipe less lightweight than the formal baseline. For a
final submission, components should therefore be selected by considering
both their AUC/LogLoss contribution and their measured inference or
maintenance cost.

This study is limited to the offline challenge setting. Unlike
production-oriented ranking papers that can report online A/B tests, this
work focuses on validation and leaderboard-style evaluation. The
discussion should therefore be read as evidence about robust offline
modeling under the released data protocol, not as a direct claim about
online business metrics.

\section{Conclusion}

We presented STAR, a structured-tokenization and target-aware
interest-representation framework for PCVR prediction in the KDD Cup
2026 Tencent UniRec Challenge. The current experiments show that the
optimized architecture benefits from moderate capacity scaling, while
the main ablation evidence highlights temporal context, contrastive
alignment, DIN-style interest encoding, and high-cardinality sequence
recovery as the most reliable modeling contributors. Future work
includes more generalized model architectures—such as UniFormer and
OneTrans—that unify feature interaction and sequence modeling, while
also investigating effective scaling strategies that ensure features are
utilized efficiently.

\section*{Ethics and Privacy Statement}

This work studies PCVR prediction using anonymized recommendation data.
The proposed method does not require personally identifiable information
and is intended for ranking-quality improvement under the privacy and
data-use constraints of the challenge. Potential fairness and feedback
loop effects should be monitored before production deployment.

\bibliographystyle{ACM-Reference-Format}
\bibliography{sample-base}
\appendix

\newpage
\section*{Appendix}

\section{Additional Implementation Details}

The full optimized model is launched with
\texttt{bash best/train/run.sh}. The default script uses RankMixer
tokenization, $d_{\mathrm{model}}=128$, sparse embedding dimension $64$,
two HyFormer blocks, two query tokens per sequence, focal loss, bf16
mixed precision, dense-parameter EMA, InfoNCE temperature $0.07$, and a
fixed global batch size of $5880$. The default sequence truncation
lengths are \texttt{seq\_a}/\texttt{seq\_b}: 256 and
\texttt{seq\_c}/\texttt{seq\_d}: 512. Ablations are run by changing only
the removed component shown in the main and appendix ablation tables.

\section{Additional Sparse-Scaling Analysis}
\label{sec:additional_scaling}

\begin{figure}[H]
  \centering
  \includegraphics[width=\columnwidth]{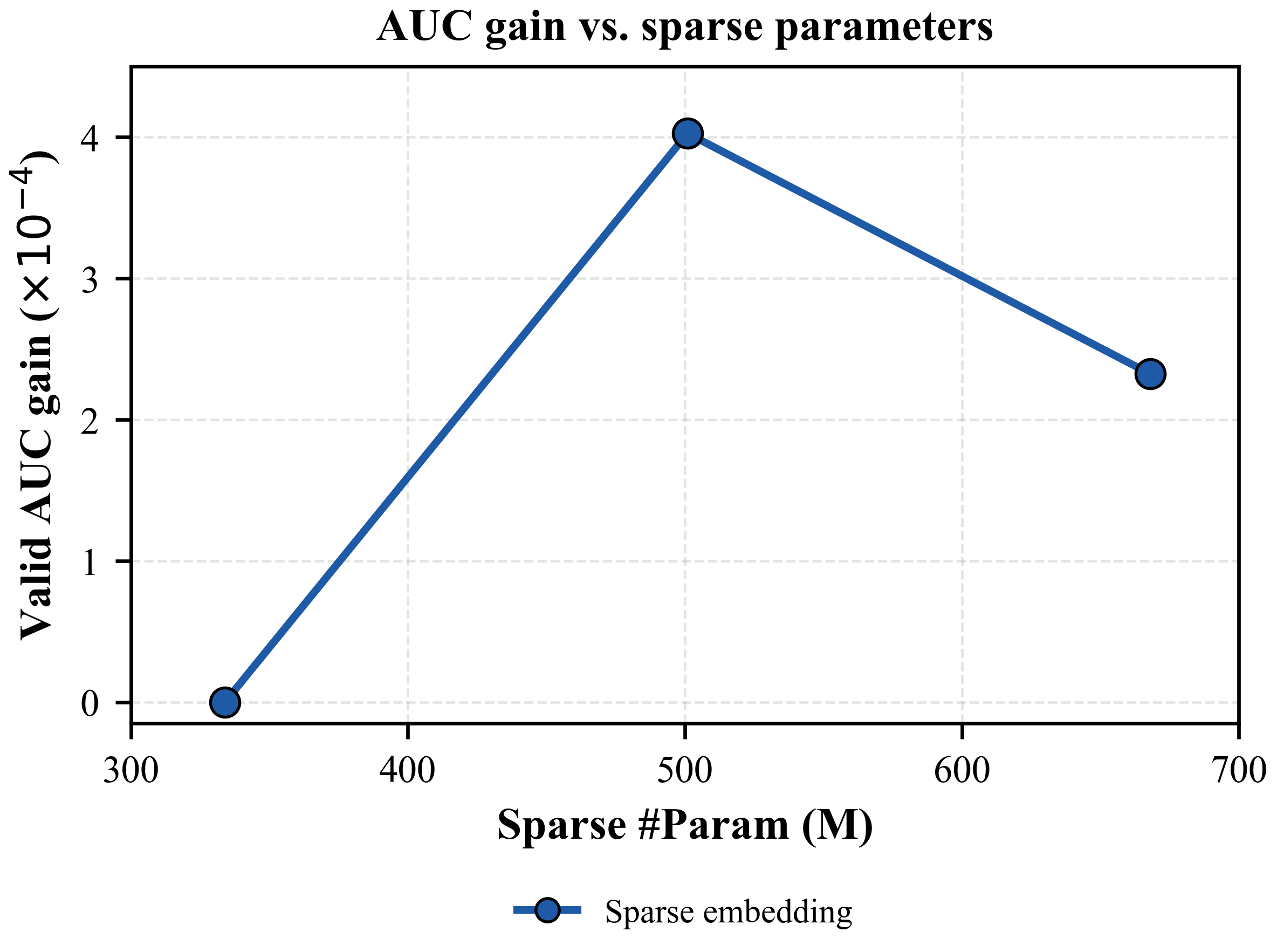}
  \caption{Validation-AUC scaling curve over sparse embedding parameters.
  The last point also changes $d_{\mathrm{model}}$ from 128 to 192, so it
  should be interpreted as a mixed sparse-and-dense capacity setting.}
  \label{fig:sparse_scaling_curves}
\end{figure}

Figure~\ref{fig:sparse_scaling_curves} summarizes the embedding-size
sweep, which shows a less monotonic pattern on validation. The sparse
parameter count increases from 333.97M to 500.96M and then 667.94M,
while validation AUC changes from 0.844503 to 0.844906 and then
0.844736. This suggests diminishing returns from simply enlarging sparse
embedding tables. Test AUC does not show the same drop, but the largest
embedding setting also changes $d_{\mathrm{model}}$ from 128 to 192, so
the test gain cannot be cleanly attributed to sparse-parameter scaling
alone.

\section{Additional Ablations}
\label{sec:additional_ablations}

Table~\ref{tab:mixed_model_ablation} reports modeling ablations whose
removal does not hurt test AUC in the current single-run setting. These
modules are therefore not included in the main ablation table. They
remain useful diagnostics because they indicate which additions may be
redundant, seed-sensitive, or over-specialized to the validation split.

\begin{table}[H]
  \centering
  \caption{Mixed modeling ablations moved out of the main table.}
  \label{tab:mixed_model_ablation}
  \scriptsize
  \setlength{\tabcolsep}{1.5pt}
  \begin{tabular}{@{}p{0.17\columnwidth}p{0.31\columnwidth}rrrr@{}}
    \toprule
    Variant & Removed or changed component
    & Val AUC & LogLoss & Test AUC & $\Delta$ Test AUC \\
    \midrule
    Full model & None
    & 0.844503 & 0.296215 & 0.836546 & 0.000000 \\
    w/o DIN token & Fused target-aware interest token
    & 0.844496 & 0.295950 & 0.836587 & +0.000041 \\
    w/o pair pooling & Weighted sparse-dense pooling for aligned pair fields
    & 0.844724 & 0.296050 & 0.836651 & +0.000105 \\
    w/o drop 89--91 & Keep user integer fields 89--91 and remove their missing residual
    & 0.844632 & 0.295822 & 0.836662 & +0.000116 \\
    w/o meta context & Disable stage-2 meta summary and stage-3 query conditioning
    & 0.844516 & 0.296400 & 0.836650 & +0.000104 \\
    \bottomrule
  \end{tabular}
\end{table}

Table~\ref{tab:protocol_ablation} reports training-protocol ablations.
These results are important for reproducing the final system but are kept
separate from the main modeling table because they evaluate optimization
and data-ordering choices rather than architectural components.

\begin{table}[H]
  \centering
  \caption{Training-protocol ablations.}
  \label{tab:protocol_ablation}
  \scriptsize
  \setlength{\tabcolsep}{1.5pt}
  \begin{tabular}{@{}p{0.17\columnwidth}p{0.31\columnwidth}rrrr@{}}
    \toprule
    Variant & Removed or changed component
    & Val AUC & LogLoss & Test AUC & $\Delta$ Test AUC \\
    \midrule
    Full model & None
    & 0.844503 & 0.296215 & 0.836546 & 0.000000 \\
    w/o EMA & Disable dense-parameter EMA checkpoint selection
    & 0.842331 & 0.290766 & 0.834403 & -0.002143 \\
    w/o shuffle & Remove the shuffle strategy
    & 0.843950 & 0.297668 & 0.835849 & -0.000697 \\
    BCE loss & Replace focal loss with BCE
    & 0.844348 & 0.212041 & 0.836411 & -0.000135 \\
    \bottomrule
  \end{tabular}
\end{table}

\section{Failed Attempts and Analysis}

In this section, we will discuss the modifications that were effective in the preliminary round but ineffective in the final round, as well as some individual modifications that were effective in both rounds but could not be combined. 
\subsection{Modifications Effective In The Preliminary Round But Ineffective In The Final Round}
The first part concerns modifications effective in the preliminary round but ineffective in the final round. 

This primarily involves adversarial training with FSGM, which showed some improvement in the preliminary round. This was partly due to the smaller size of the preliminary dataset compared to the final round, posing a risk of overfitting. Adversarial training effectively mitigated this problem. However, in the final round, FSGM could not be consistently integrated into the main solution.

Regarding temporal features, the strategy used in the preliminary round could not be effectively integrated; this was because the preliminary round's approach incorporated a discrete "day-of-week" feature, and the resulting embedding introduced noise during the semi-final round due to differences in the test data distribution.

In addition, we observed a clear data-scale-dependent preference in the choice of sequence encoder. In the preliminary stage, the Transformer-based sequence encoder performed better, whereas in the final stage, the attention-free SwiGLU encoder achieved superior performance. Since the amount of training data in the preliminary stage was only about \verb|1/19| of that in the final stage, this result suggests that the effectiveness of a sequence encoder depends not only on its modeling capacity, but also on the available data scale and computational budget. This phenomenon indicates a data-scale-driven architectural transition: in the small-data regime, Transformer benefits from its stronger sequential inductive bias; in the large-data regime, SwiGLU becomes preferable due to its efficiency, implicit regularization effect, and clearer division of labor with downstream target-aware interaction modules.
\subsection{Partially Effective But Difficult To Combine}
The second part consists of changes that are partially effective, but difficult to integrate. These include changing the dense compression from a single token to 2-5 tokens and the utilization of the item id. 

Expanding the dense representation from a single token to multiple tokens works on the baseline individually, but fails in the final ensemble. This expansion likely disrupts the compression bottleneck, increasing the risk of overfitting.

Using item IDs, especially with unknown tokens, is only effective under certain dropout ratios during training; it may have negative benefits under other hyperparameters.
\subsection{Better Along The Way, Worse In The Final Solution}
There were improvements along the way, but points were lost when the final solution could not be achieved. 

On the one hand, calculate an additional global Target-aware interest token and concatenate it to the NS Tokens. Because DINMLPQueryDecoder already gives all Sequence Queries strong Target-aware capabilities, adding an extra global DIN Token introduces information redundancy, potentially diluting the attention of other features in the RankMixer Block.

On the other hand, extract meta features such as sequence length, truncation rate, and time span and add them to the prediction. It is prone to overfitting to specific splits of the validation set. In scenarios where there is a distribution shift in train-inference, meta features are often most susceptible to shift, leading to dropped test set data.
\subsection{Other Failed Attempts Summary}
In addition, we conducted numerous experiments, trying various methods for generating queries using different windows and long sequence modeling (e.g., longer), but without achieving significant gains. These results are related to the distribution of the data and the regularization strength of the proposed solution.

Other interesting points include the fact that removing some invalid features may decrease the value but increase the test result, which could be due to reducing the likelihood of overfitting.

An important observation during the competition was that while certain architectural changes appeared locally optimal at specific stages, they were not globally optimal in the final ensemble.

Another key finding is that the benefits of scaling up rely on proper feature handling and a sufficient volume of data. Our attempts to scale up during the competition were premature; at that stage, we had not adequately addressed feature processing, which made the models highly prone to overfitting and resulted in significant deficiencies in how capacity was allocated to utilize each effective feature. Consequently, our approach failed to effectively scale up while simultaneously resolving the issues of feature processing and utilization.

\end{document}